\documentclass[prl,twocolumn,showpacs,preprintnumbers]{revtex4}
\usepackage{graphicx}
\usepackage{amsmath}
\usepackage{amssymb}
\usepackage{times}

\def\b{\beta}

\def\d{\delta}
\def\e{\epsilon}

\def\S{\Sigma}
\def\s{\sigma}

\def\w{\omega}
\def\ua{\uparrow}
\def\da{\downarrow}

\def\Vec#1{\mathbf #1}
\begin{document}

\title{Evolution of electronic structure of doped Mott insulators \\
- reconstruction of poles and zeros of Green's function }

\author{Shiro Sakai, Yukitoshi Motome, and Masatoshi Imada}

\affiliation{Department of Applied Physics, University of Tokyo, Hongo,
Tokyo 113-8656, Japan}

\date{\today}

\begin{abstract}
We study evolution of metals from Mott insulators in the carrier-doped 
2D Hubbard model using a cluster extension of the dynamical mean-field theory.
While the conventional metal is simply characterized 
by the Fermi surface (pole of the Green function $G$), 
interference of the {\it zero surfaces} of $G$ with the pole surfaces becomes 
crucial in the doped Mott insulators. 
Mutually interfering pole and zero surfaces are dramatically transferred over 
the Mott gap,
when lightly doped holes synergetically loosen the doublon-holon binding.
The heart of the Mott physics such as the pseudogap, hole pockets, Fermi arcs, 
in-gap states, 
and Lifshitz transitions appears as natural consequences of this global interference in the frequency space.  
\end{abstract}
\maketitle


Electronic structures of doped Mott insulators have intensively been debated since the discovery of high-$T_{\text c}$ cuprates.
While overdoped metals behave like 
a conventional Fermi liquid,
an outstanding issue is how such metals evolve into 
the Mott insulator.
Experimental studies on the doped cuprates revealed
anomalous metallic behaviors in the proximity to the Mott insulator.
Among them angle-resolved photoemission spectroscopy (ARPES)
found arclike spectra around the nodal points of the momenta 
$\Vec{k}=(\pm\frac{\pi}{2},\pm\frac{\pi}{2})$, 
with a pseudogap in the antinodal region
of the Brillouin zone \cite{nd98}.
Moreover the high-resolution ARPES \cite{cs08} and quantum 
oscillations of resistivity observed under magnetic fields~\cite{dp07}
suggested the existence of a small closed Fermi surface, 
called the pocket, in contrast to the 
large surface expected in the normal Fermi liquid.
These imply emergence of non Fermi liquids under a radical evolution of electronic structure upon doping.

Diverse theoretical proposals were made for
the Mott physics of the 2D Hubbard model:
The pseudogap was reproduced in various theoretical frameworks \cite{yj03,ph95,kk06}. 
The exact diagonalization (ED) studies \cite{do92} 
further suggested a spectral weight transfer 
with doping from the upper Hubbard band (UHB) to the upper edge of 
the lower Hubbard band (LHB), which created in-gap states.
The arc structure of the spectral weight 
was reproduced in the cluster perturbation theory \cite{st04},
the dynamical cluster approximation \cite{mj06,gw08},
and the cellular dynamical mean-field theory (CDMFT) 
\cite{cc05,kk06,sk06,hk07}.
Hole-pocket Fermi surfaces around 
the nodal points were suggested in a phenomenology \cite{kr06},
the CDMFT \cite{sk06},
and a variational-cluster approach \cite{aa06}.
A topological change
of the Fermi surface, i.e., Lifshitz transition \cite{l60},
from electronlike to holelike surfaces due to a correlation effect was analyzed
\cite{mp02,kf05}.
A Lifshitz transition to electron pockets 
was also found upon electron doping \cite{hi06}.
In spite of these achievements, a coherent picture of the Mott physics has not emerged yet.

In general, poles of the single-particle Green function 
$G(\Vec{k},\omega)$ 
are known to define the Fermi surface at the frequency $\omega=0$.  
While Re $G$ changes its sign across a pole through $\pm \infty$, Re $G$ may also change the sign across a {\it zero} defined by $G=0$.
Recent studies suggest, in addition to the poles, zeros of 
$G$ play important roles \cite{d03,kr06,sk06}.
In the Mott insulator, the self-energy $\S(\Vec{k},\omega)$ inside the gap
diverges on a specific surface in the $\Vec{k}$-$\w$ space, 
which defines a zero surface of $G$. Since the Fermi surface disappears in the Mott insulator while the zero surface appears instead,
the evolution of the zeros makes crucial effects at low doping and is imperative in understanding the Mott physics.

In this letter we study the 2D Hubbard model using the 
CDMFT+ED method \cite{sk06}, and clarify the doping evolution of 
poles and zeros of $G$ in the $\Vec{k}$-$\omega$ space.
By figuring out reconstruction of interfering poles and zeros
in a wide $\omega$ range, fragmentary features are integrated 
into a coherent understanding of the Mott physics
at and around $\omega=0$, namely,
the hole-pocket Fermi surface, Fermi arc, pseudogap,
in-gap states, and Lifshitz transitions. 
Starting from a high-energy structure
in the scale of the Mott gap, we zoom in lower-energy
hierarchy, in a clear and visible fashion, 
revealing two structures induced by hole doping.
One is in-gap states formed above the occupied states separated 
by a pseudogap. 
The other is hole pockets in the nodal directions, 
which appear at the top of the occupied states below the pseudogap.  
The pseudogap is described by a zero surface crossing the
Fermi level.
The density-of-states analysis suggests that these structures result from a 
dramatic transfer of the weight over the Mott gap:
This is caused by tiny doping, which loosens the doublon-holon binding.
Then the zero surface interferes with the poles and pushes down 
the pole surface near the zeros below the Fermi level, 
leaving hole pockets 
around ($\pm\frac{\pi}{2},\pm\frac{\pi}{2}$), which are 
more stabilized by the next-nearest transfer.
With further doping, the Fermi surfaces undergo at least
two continuous Lifshitz transitions before reaching a normal
Fermi liquid with an electronlike Fermi surface. 


The Hubbard Hamiltonian on a square lattice reads
\begin{eqnarray}
  H= \sum_{\Vec{k}\s}\e(\Vec{k})c_{\Vec{k}\s}^\dagger c_{\Vec{k}\s}
   -\mu \sum_{i\s}n_{i\s}+ U\sum_{i\s}n_{i\ua}n_{i\da},
  \label{eq:hubbard}
\end{eqnarray}
where $c_{\Vec{k}\s}$ $(c_{\Vec{k}\s}^\dagger)$ destroys (creates) 
an electron of spin $\s$ with momentum $\Vec{k}$, and 
$n_{i\s}$ is a spin density operator at site $i$.
$U$ represents the onsite Coulomb repulsion, $\mu$ the chemical
potential, and 
$\e(\Vec{k})=-2t(\cos k_x + \cos k_y) -4t' \cos k_x\cos k_y$,
where $t$ $(t')$ is the (next-)nearest-neighbor transfer integral.

\begin{figure}[t]
  \center{
  \includegraphics[width=7.5cm]{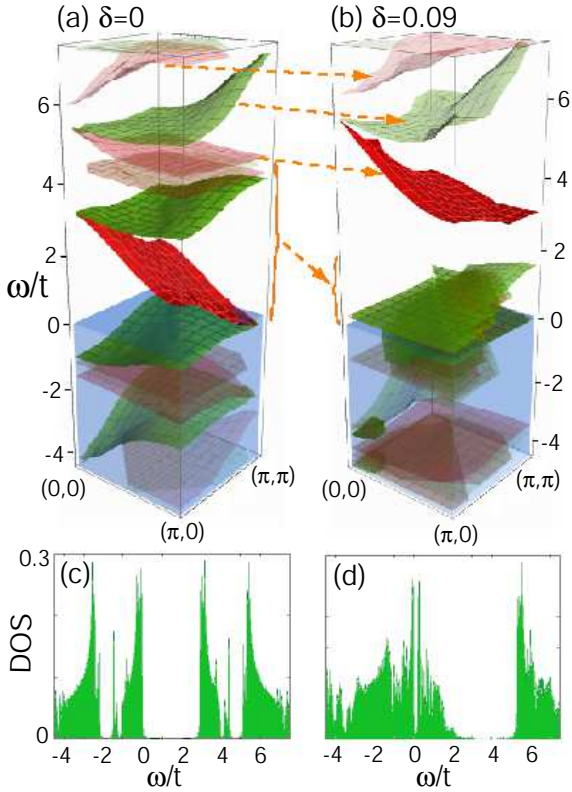}}
  \caption{The $\Vec{k}$-$\w$ structure of poles and 
           zeros of $G$ for $t'=0$, $U=8t$ at
           (a) $\d$=0 and (b) 0.09. 
           Poles (zeros) are plotted in green (red).
           Occupied regions for an electron are filled with aqua.
           The transparency of pole (zero) surfaces reflects 
           the imaginary part of the (inverse) self-energy
           (more transparent for larger imaginary part).
           (c) and (d) show the density of states per spin
           at $\d=0$ and 0.09, respectively.}
  \label{fig:r3d}
\end{figure}

In the CDMFT, the infinite lattice of the model (\ref{eq:hubbard})
is self-consistently mapped onto an $N_{\text c}$-site cluster 
embedded in infinite bath sites, 
which define an $N_{\text c} \times N_{\text c}$ dynamical 
mean-field matrix ${g}_0(i\w_n)$ for the cluster in terms of 
Matsubara frequency $\w_n$.
We employ the ED method with the Lanczos algorithm to solve
the cluster problem at zero temperature, where 
${g}_0$ is fitted with a finite number of bath parameters and a fictitious
temperature $1/\b$ is introduced.
We adopt 2 by 2 cluster ($N_{\text c}=4$) coupled 
with 8 bath sites and $\b=100/t$ unless otherwise mentioned.
When a self-consistency loop converges, 
we obtain an $N_{\text c} \times N_{\text c}$ 
self-energy matrix $\S_{\text c}(i\w_n)$,
whose elements are defined on sites within the cluster.
Then we interpolate cluster quantities in the momentum space,
to obtain the original infinite-lattice ones.
To study poles and zeros of $G$ simultaneously, 
after careful examination, we employ a suitable scheme of the 
cumulant periodization~\cite{sk06}, 
where the cluster cumulant 
$M_{\text c} \equiv(i\w_n+\mu-\S_{\text c})^{-1}$ is interpolated
with a Fourier transformation cut by a cluster size $N_{\text c}$,
\begin{eqnarray}\label{eq:periodize}
  M(\Vec{k},i\w_n)=\frac{1}{N_{\text c}}\sum_{i,j=1}^{N_{\text c}}
                    [M_{\text c}(i\w_n)]_{ij}
                    e^{i\Vec{k}\cdot(\Vec{x}_i-\Vec{x}_j)}.
\end{eqnarray}
After $M$ is obtained, $G$ and $\S$ on the original lattice
are calculated with $G =[M^{-1}-\e(\Vec{k})]^{-1}$ and 
$\S= i\w_n+\mu-M^{-1}$, respectively.
The Lanczos-ED method allows us to calculate 
$G$ at real frequencies directly through a 
continued-fraction expansion \cite{gb87},
where a Matsubara frequency $i\w_n$
is replaced by $\w+i\eta$ with 
a small positive factor $\eta$.
Although we should be careful on the limitation of the present size 
of the cluster, this combination of the methods is appropriate as 
the state of the art in exploiting zero and pole structures on equal 
footing in a wide spectrum range. 



\begin{figure}[t]
  \center{
  \includegraphics[width=7.5cm]{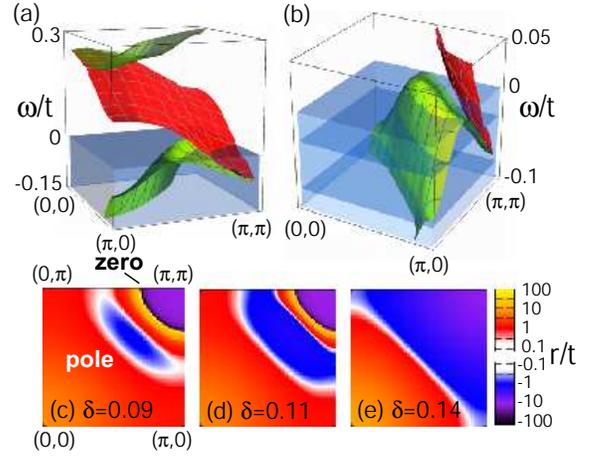}}
  \caption{(a)(b) Close-ups of Fig.\ \ref{fig:r3d}(b)
           at two different lower-energy scales. 
           Blue grade levels in (b) depict rigid-band shifts of 
           the chemical potential.
           (c)(d)(e) 
           $r(\Vec{k})\equiv {\rm Re}[G^{-1}(\Vec{k},0)]$ 
           in the first quadrant of the Brillouin 
           zone for $t'=0$ and $U=8t$ at $\d=$0.09, 0.11 and 0.14, 
           respectively.
           $r=0$ ($\pm\infty$) defines the Fermi surface
           (the zero surface of $G$).
           We set $\eta=10^{-4}t$ for these data.}
  \label{fig:pocket}
\end{figure}

We start from electronic structure of the Mott insulator at half filling 
($\delta=0$; $\delta$ is the hole doping rate from half filling)
for $t'=0$ and $U/t=8$.
While an insulating phase caused by some symmetry breaking 
may show a similar feature, 
to capture the essence of the Mott physics, 
we assume a paramagnetic solution in the CDMFT. 
Figure\ \ref{fig:r3d}(a) shows the structure of poles (green surfaces) and zeros (red surfaces) of $G$ in the $\Vec{k}$-$\omega$ space.
Here $\omega$ is measured from the upper edge of the 
occupied states~\cite{note}
to compare with the doped case below.
The zero surface around $0 < \omega < 3t$
(the most distinct red surface) represents 
the Mott gap, which separates UHB and LHB displayed by entangled pole and zero surfaces.  
The zero surface at the Mott gap touches the pole surfaces 
of the UHB (LHB) around (0,0) [($\pi,\pi$)] at $\w \simeq 3t$ ($\w\simeq 0$).
The distinct pole surface in the UHB persists up to $\w\sim 4t$, and 
another distinct one appears around $5t < \omega < 7t$.
In accordance with these pole surfaces,
there are large weights in the density of states (DOS) per spin
[Fig.~\ref{fig:r3d}(c)].
In other area, poles are disrupted due to a large imaginary part of 
the self-energy, making incoherent contributions.


We next dope holes.
Comparing the result at $\d=0.09$ in Fig.\ \ref{fig:r3d}(b) with
Fig.\ \ref{fig:r3d}(a), we see that 
high-energy ($|\w|\gtrsim 5t$) structure does not change
appreciably while the lower-energy ($|\w|\lesssim 5t$) structure does: 
The pole surface at the bottom of UHB just above 
the Mott gap $\sim 3t$ in Fig.\ \ref{fig:r3d}(a) is transferred to 
the lowest-energy pole surface in the piles above $\w \simeq 0.2t$,
squashing the zero surface just below it, as in Fig.\ \ref{fig:r3d}(b) and
as will be seen in Figs.~\ref{fig:pocket}(a) and (b).
In addition, one of vague zero surfaces in the UHB piles 
in Fig.\ \ref{fig:r3d}(a) turns to 
a distinct one at around $\w=4t$ in Fig.\ \ref{fig:r3d}(b).
Corresponding to these reconstructions,
the DOS at $\w \simeq 3t$ in Fig.~\ref{fig:r3d}(c) is transferred 
to energies lower than $\w \simeq 2t$ as shown in Fig.~\ref{fig:r3d}(d).

Let us focus on a low-energy structure of poles and zeros.
Figures\ \ref{fig:pocket}(a) and (b) provide enlarged views of 
Fig.\ \ref{fig:r3d}(b) near the Fermi level,
in two different energy scales.
A prominent feature is a small gap, described by a zero surface
cutting the Fermi level.
The gap has a small width of $\sim 0.2t$, and is
distinguished from the larger gap between
the doped hole states and the UHB
as shown in DOS in Fig.~\ref{fig:r3d}(d).
A comparable pseudogap was 
found in earlier CDMFT studies \cite{sk06,kk06}.

As the zero surface around ($\pi,\pi$) extends to the region $\w <0$,
near the zero surface, the pole surface is pushed down below the Fermi 
level [Fig.~\ref{fig:pocket}(b)], 
because a zero and a pole surfaces cannot cross each other.
Meanwhile in the regions far away from zeros,
the energy of poles increases with $|\Vec{k}|$, reflecting 
the original dispersion $\e(\Vec{k})$.
Hence, along the direction from (0,0) to ($\pi,\pi$) for example,
the pole energy initially increases, crossing the Fermi level,
and then turns down around ($\frac{\pi}{2},\frac{\pi}{2}$), 
crossing the Fermi level again.
As a consequence, a hole pocket is formed 
around ($\frac{\pi}{2},\frac{\pi}{2}$), accompanying a zero surface
around ($\pi,\pi$), as found previously in Ref.\ \onlinecite{sk06}
and also shown in Fig.\ \ref{fig:pocket}(c).

The metal-insulator transition occurs when 
the hole pockets shrink and vanish, indicating a topological
nature of this transition~\cite{misawa}.
Since the quick evolution for $\delta\lesssim 0.01$ elucidated below 
suggests a proximity to the first-order transition,
the marginal quantum criticality~\cite{misawa} may have relevance.

\begin{figure}[t]
  \center{
  \includegraphics[width=0.47\textwidth]{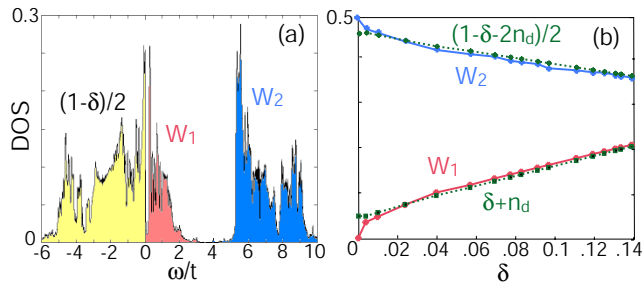}}
  \caption{(Color online) (a) DOS at $\d=0.09$ 
          [the same as Fig.~\ref{fig:r3d}(d)]
           and the definitions of weights $W_1$ and $W_2$.
           (b) $\d$ dependence of $W_1$ and $W_2$
           compared with the weights estimated from a simple analysis
		   given by dashed lines (see the text).
           We set $\eta=10^{-2}t$ for these data.}
  \label{fig:DOS}
\end{figure}

To get further insight into the underlying physics of the dramatic 
restructuring, we study the doping evolution of DOS systematically. 
We calculate the weight integrated over energy ranges of
$0<\w<2.8t$ (in-gap weight), $W_1$, and that over $2.8t<\w<\infty$ 
(UHB weight), $W_2$ [see Fig.~\ref{fig:DOS}(a)].
Figure \ref{fig:DOS}(b) depicts $\d$ dependences of $W_1$ and $W_2$,
where $W_1$ $(W_2)$ monotonically increases (decreases) with $\d$.

As shown in Fig.~\ref{fig:DOS}(b), the in-gap weight $W_1$ shows
good agreement with $\delta + n_{\text d}$, 
where $n_{\text d}\equiv \langle n_{i\uparrow}n_{i\downarrow}\rangle$
is the doublon density.
It is natural to have the weight proportional to $\delta$ 
in the rigid band picture because it 
comes from the holes doped into the LHB,
but there is an additional contribution scaled by $n_{\text d}$.
This is interpreted as a strong-correlation effect as follows.
Neglecting dynamical fluctuations, we can roughly estimate
the single-occupancy weight as 
$W_0\equiv \frac{1}{2}\sum_\s (n_\s-n_{\text d})
=\frac{1}{2}(1-\d-2n_{\text d})$,
where $n_\s \equiv \langle n_{i\s} \rangle$ ($\s=\ua,\da$) is 
the spin density.
Now, one electron added to these singly occupied states makes 
a ``doublon" and contributes to the DOS of the UHB. 
This contribution should agree with $W_2$.
Since the DOS below the Fermi level has a weight 
$\frac{1}{2}(1-\d)$, the remaining weight, 
which should be equal to $W_1$, becomes 
$1-\frac{1}{2}(1-\d-2n_{\text d})-\frac{1}{2}(1-\d)=\d+n_{\text d}$.
This simple analysis explains well the overall behavior 
of $W_1$ and $W_2$, as seen in Fig.~\ref{fig:DOS}(b).

A nontrivial finding, however, is that 
the weight $n_{\text d}$ in $W_1$ is very quickly transferred from the UHB 
to the top of LHB by tiny doping $\delta \lesssim 0.01$ to the Mott insulator.
Actually, the above simple analysis
breaks down at very small $\delta$, where 
$W_1$ eventually and obviously vanishes toward $\delta=0$ 
as shown in Fig.~\ref{fig:DOS}(b).  
At $\delta=0$, this weight $n_{\text d}$ is included in 
the UHB weight $W_2$ and is interpreted as an electron excitation added 
to a holon tightly bound with a doublon.
The number of doublons and holons should be the same 
in the Mott insulator and they are tightly bound with the 
binding energy equal to the Mott gap $\sim 3t$. 
Therefore one electron added to a hole site requires dissolution 
of the bound state and overcoming the Mott gap. 
The quick transfer of $n_{\text d}$ from $W_2$ to $W_1$ means that 
the doublon-holon binding energy is quickly reduced from $\sim 3t$ to 
$\sim 0.2t$
presumably because of an efficient screening of 
the doublon-holon interaction by the doped holes. 
Even with this screening, the binding energy though small still 
survives as the pseudogap.
This quick reduction is promoted by a positive feedback, because the 
dissolved bound pairs further join in screening other bound pairs. 
This constitutes a mechanism for the drastic restructuring in the Mott physics.

The next issue is how the hole-pocket Fermi surface accompanied 
by zeros of $G$ in Fig.~\ref{fig:pocket}(c)
evolves into a single electronlike Fermi 
surface expected in the normal Fermi liquid as $\d$ increases further.
Figures \ref{fig:pocket}(c), (d), and (e) depict 
$r(\Vec{k})\equiv {\rm Re}[G^{-1}(\Vec{k},0)]$ 
at $\d=0.09$, 0.11, and 0.14, respectively, showing how Fermi (zero)
surfaces at $r=0$ $(\pm\infty)$ evolve.
As $\d$ increases from 0.09, 
the hole pocket continues to expand until it touches the Brillouin 
zone boundary ($|k_x|=\pi$ or $|k_y|=\pi$).
When it touches the boundary, it changes into two concentric 
Fermi surfaces around ($\pi,\pi$) through a Lifshitz transition,
while the zero surface remains around ($\pi,\pi$) [Fig.\ \ref{fig:pocket}(d)].
Further doping enlarges the unoccupied region 
(blue region) sandwiched by the two concentric Fermi surfaces.
Then, the smaller Fermi surface around ($\pi,\pi$) merges with 
the zero surface and they are annihilated in pair, 
leaving a large holelike Fermi surface, 
which almost simultaneously transforms into a normal electronlike 
one through another Lifshitz transition [Fig.\ \ref{fig:pocket}(e)].
For $\d\geq 0.14$ only a large electronlike Fermi surface is found.
This evolution of Fermi and zero surfaces 
is understood in a rigid band picture for the electronic structure
in Fig.\ \ref{fig:pocket}(b), 
as drawn with blue grade levels,
if we assume that the hole doping only lowers the Fermi level 
without changing the structure of poles and zeros.
Detailed inspection of the two Lifshitz transitions shows 
continuities of the electron density as a function of $\mu$ and hence 
continuous transitions.

Apparently, the Luttinger sum rule \cite{l60-2} is 
violated for $\d\lesssim 0.11$ while roughly 
satisfied for $\d\gtrsim 0.14$.
Since the Luttinger theorem assumes an adiabatic continuity, there is no reason that the rule should 
be satisfied beyond the Lifshitz transition around
$\d=0.14$.
Other numerical calculations 
also pointed out the violation at small $\delta$
\cite{sk06,mp02,kf05}. 

Next we consider effects of $t'$.
Results for $\d=0.09$ (not shown) exhibits that 
$t'<0$ elevates (lowers) the energy of poles around the nodal 
direction [around $(\pi,0)$ and $(0,\pi)$],
as expected from the original dispersion $\e(\Vec{k})$.
The enhancement around the nodal direction more stabilizes
the hole-pocket structure while the reduction around $(\pi,0)$ and
$(0,\pi)$ more stabilizes a single holelike Fermi surface. 
Quantitative modifications caused by nonzero $t'$ will be reported elsewhere.

\begin{figure}[t]
  \center{
  \includegraphics[width=0.35\textwidth]{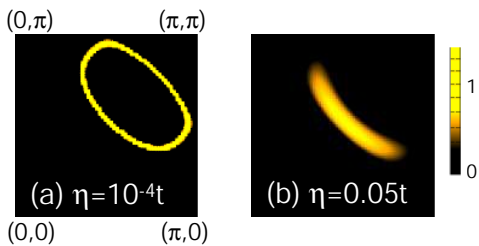}}
  \caption{(Color online) $A(\Vec{k},0) \equiv -$Im$[G(\Vec{k},\w+i\eta)]/\pi$ 
           for $t'=-0.2t$, $U=12t$, and $\d=0.09$
           with (a) $\eta=10^{-4}t$ and (b) $\eta=0.05t$.
           Here we set the fictitious temperature $1/\b=t/200$.}
  \label{fig:arc}
\end{figure}

Finally we discuss how a Fermi arc emerges in the lightly doped region.
Stanescu and Kotliar found that the zero surface reduces the spectral weight
from the neighboring Fermi surface, resulting in a Fermi arc~\cite{sk06}.
In the present Lanczos-ED method, however,
one has to take the limit of $\eta \to 0^+$ in principle,
which leads to $\delta$-function peaks in Im$\Sigma$ at 
the zero surfaces of $G$.
In this limit there is no contribution from the zeros to the Fermi 
surface region,
namely, the closed hole-pocket structure remains robust in 
the spectral weight and the Fermi arc does not show up, 
as demonstrated in Fig.~\ref{fig:arc}(a) for a sufficiently small $\eta$.
In real situations of experiments, however, Im$\S$ is broadened 
through various extrinsic factors, such as
temperature, impurity scatterings and phonons
\cite{footnote}.
These effects are phenomenologically represented by a broadening to Im$\S$
through a nonzero $\eta$.
In fact, for larger $\eta$, the singularities of Im$\Sigma$ are smeared out and
the spectral weight close to the zeros are suppressed as shown in Fig.~\ref{fig:arc}(b).
Thus the Fermi arc is reproduced by introducing
a phenomenological broadening factor $\eta$ in the present result.


To summarize, we have shown that key elements of the Mott physics 
such as in-gap states, hole pocket, Fermi arc,
pseudogap, and Lifshitz transitions result from global
reconstructions and interferences of pole and zero surfaces with 
doping to the Mott insulator.
The reconstructions are caused by a drastic relaxation of the 
doublon-holon binding at tiny doping.
The criticality at $\delta\rightarrow 0$ and a mechanism of 
the relaxation, though we have given a qualitative picture,
should further be clarified in future studies.
It is also desired to confirm our results in calculations
for larger clusters.


We thank Y. Z. Zhang for useful comments. SS also thanks 
S. Watanabe, Y. Yanase, G. Sangiovanni, and 
D. Tahara for valuable discussions. 
This work is supported by a Grant-in-Aid for Scientific Research on
Priority Areas ``Physics of Superclean Materials" from MEXT, Japan.


\begin{references}
\bibitem{note}
  The electronic structure is symmetric around $\w = 1.46t$ at $\delta=0$
  due to the particle-hole symmetry.
\bibitem{footnote}
  It is also possible that Im$\S$ acquires a finite width in 
  the thermodynamic limit while the present calculation,
  based on finite clusters, cannot reproduce it.
\end{references}
\end{document}